\documentclass{article}
\usepackage{ijcai21}

\usepackage{times}

\usepackage{soul}
\usepackage{url}
\usepackage[hidelinks]{hyperref}
\usepackage[utf8]{inputenc}
\usepackage[small]{caption}
\usepackage{graphicx}
\usepackage{amsmath}
\usepackage{booktabs}
\usepackage{amsthm}
\usepackage{amsfonts}
\title{Self-Adaptive Swarm System (SASS)}

\author{
Qin Yang\\
\affiliations
Heterogeneous Robotics Research (HeRo) Laboratory, University of Georgia\\
\emails
\{qy03103\}@uga.edu
}

\begin{document}

\maketitle

\begin{abstract}
Distributed artificial intelligence (DAI) studies artificial intelligence entities working together to reason, plan, solve problems, organize behaviors and strategies, make collective decisions and learn.
This Ph.D. research proposes a principled Multi-Agent Systems (MAS) cooperation framework, \textit{Self-Adaptive Swarm System (SASS)}, to bridge the fourth level automation gap between perception, communication, planning, execution, decision-making, and learning.
\end{abstract}

\section{Introduction}


In the artificial systems, distributed artificial intelligence (DAI) has developed more than three decades as a subfield of artificial intelligence (AI). It has been divided into two sub-disciplines: Distributed Problem Solving (DPS) focuses on the information management aspects of systems with several branches working together towards a common goal; Multi-Agent Systems (MAS) deals with behavior management in collections of several independent entities, or agents \cite{stone2000multiagent}. For cooperative MAS, the individual is aware of other group members, and actively shares and integrates its needs, goals, actions, plans, and strategies to achieve a common goal and benefit the entire group, especially building so-called \textit{artificial social systems} \cite{wooldridge2009introduction}, such as Multi-Robot Systems (MRS).

For low-level planning and control, \cite{rizk2019cooperative} groups the system based on the cooperative tasks' complexity as four levels of automation, and no references were found currently. For high-level MAS decision-making and learning, recent studies mainly concern partial cooperation and do not consider deeper cooperative relationships among agents representing more complex team strategies \cite{rizk2018decision}. Combining the information from perceiving the environments and inferring the corresponding strategies and the world's conditional (or even causal) relations in those scenarios, a unified probabilistic framework to tightly integrate deep learning and Bayesian models adapting the environments and achieving the tasks with reasonable time complexity and efficient and effective information exchange between the perception component and the task-specific component are still challenging problems \cite{wang2020survey}.

\begin{figure}[t]
\centering
\includegraphics[width=0.5\textwidth]{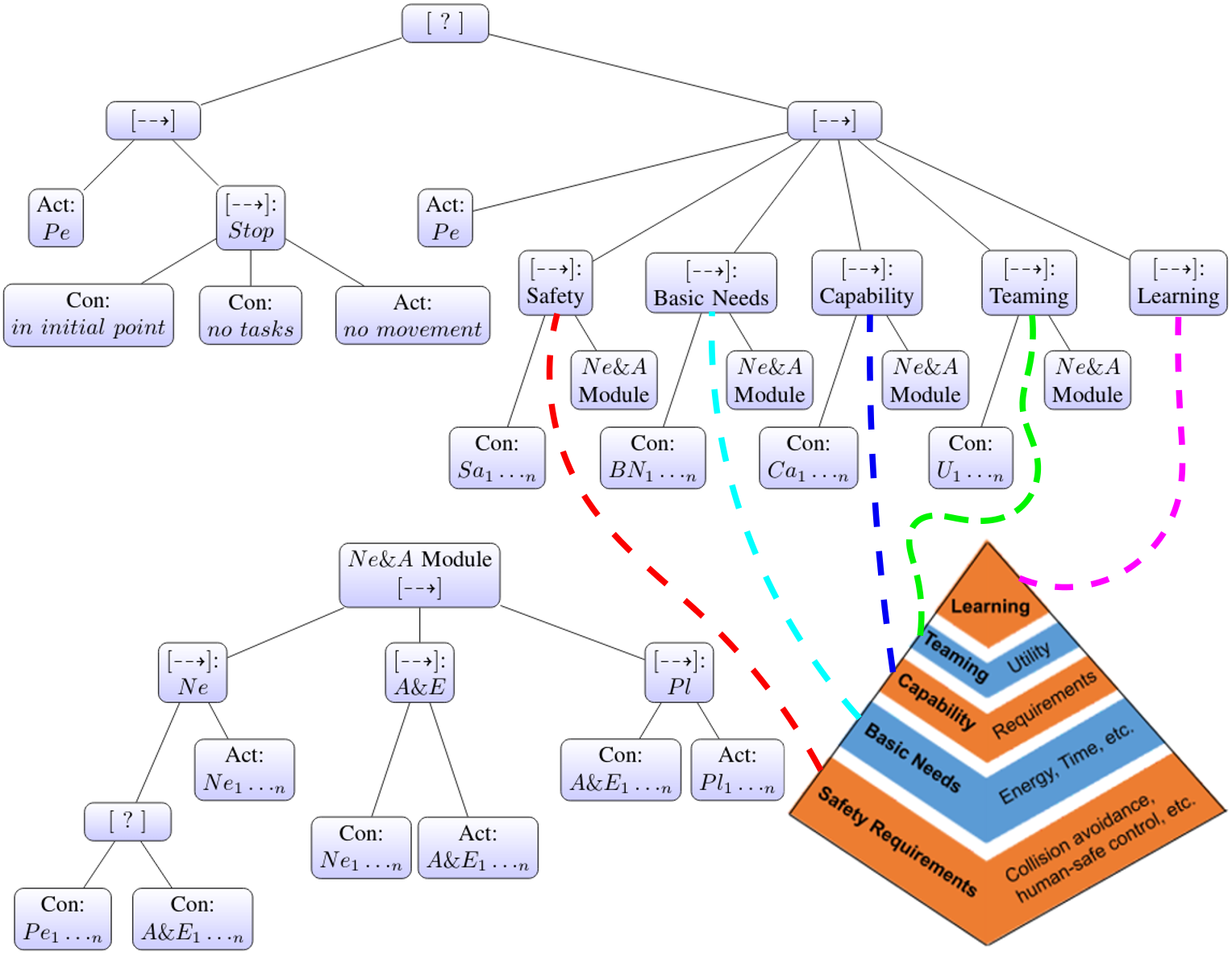}
\caption{\small Behavior Tree representing \textit{Robot Needs Hierarchy} at each agent in \textit{SASS}. $[?]$ - Selector Node, $[\dashrightarrow]$ - Sequence Node, $Con$ - Conditions, $Act$ - Actions, $Pe$ - Perception, $Sa$ - Safety, $BN$ - Basic Needs, $Ca$ - Capability, $U$ - Utility, $Pl$ - Plan, $Ne$ - Negotiation, $A\&E$ - Agreement and Execution.}
\label{fig: sass}
\end{figure}

\section{Contributions}

This \textit{Ph.D.} research proposes a principled MAS cooperation framework called \textit{Self-Adaptive Swarm System (SASS)} \cite{mrs2019,yang2020hierarchical}. 
Fig. \ref{fig: sass} shows \textit{needs-driven SASS} mechanism represented as a \textit{Behavior Tree} and we briefly introduce our contributions as follow:

\paragraph{Robot Needs Hierarchy} To model an artificial intelligence agent's motivations and needs, we classify the \textit{Robot Needs Hierarchy} as five different levels: \textit{safety needs} (collision avoidance, human-safety control, etc.); \textit{basic needs} (energy, time constraints, etc.); \textit{capability} (heterogeneity, hardware differences, etc.); \textit{teaming} (global utility, team performance, etc.); and \textit{self-upgrade} (learning).

        
\paragraph{Negotiation-Agreement Mechanism} We propose a \textit{distributed  Negotiation-Agreement} mechanism for selection (task assignment), formation (shape control), and routing (path planning) through automated planning of state-action sequences, helping MAS solve the conflicts in cooperation.

\begin{figure}[t]
\centering
\includegraphics[width=0.485\textwidth]{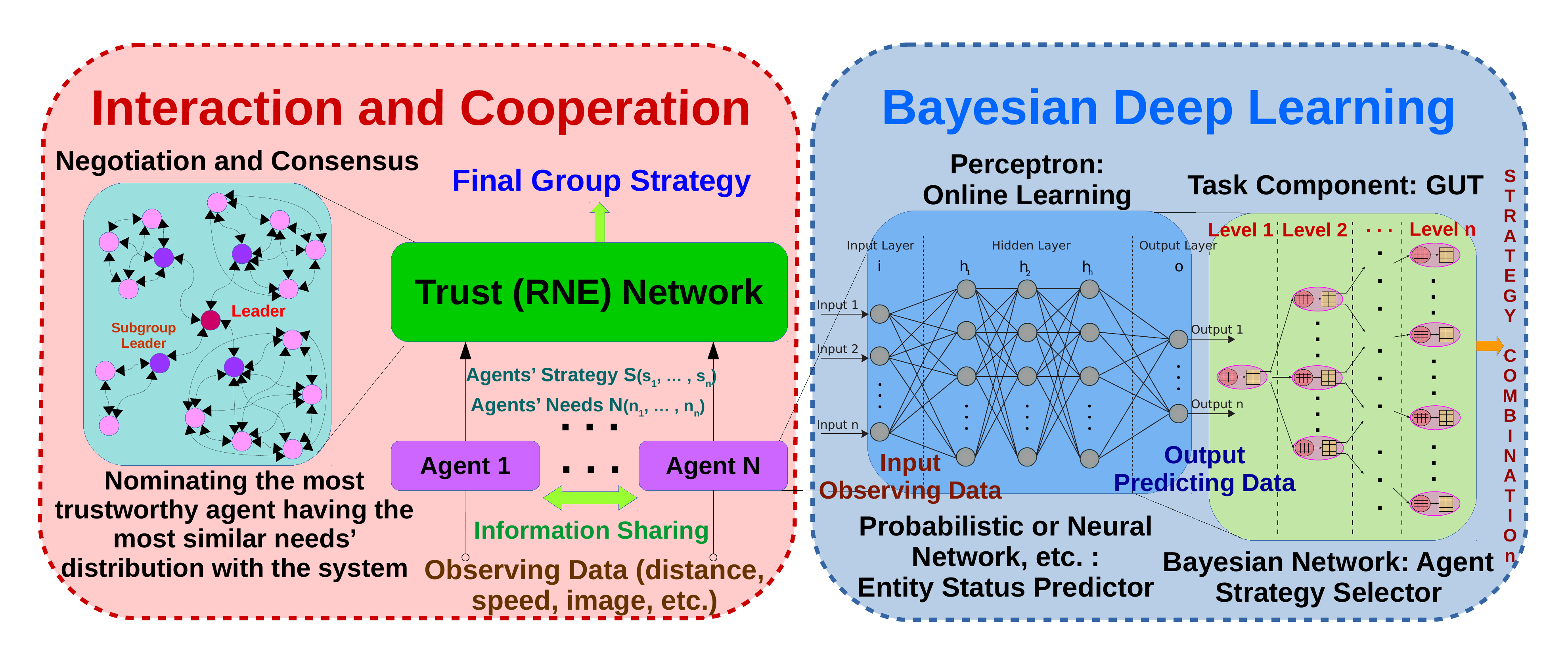}
\setlength{\abovecaptionskip}{-10pt}
\caption{\small SASS Synthetic Framework}
\vspace{-4mm}
\label{fig: sass_sf}
\end{figure}
    
\paragraph{Atomic Operations} By decomposing the complex tasks into a series of simple sub-tasks based on the \textit{Atomic Operations}: \textit{Selection}, \textit{Formation}, and \textit{Routing}, agents can recursively achieve all sub-tasks until MAS completes the mission. 
        
\paragraph{Game-theoretic Utility Tree (GUT)} \textit{GUT} \cite{yang2020game} is a new state-space Bayesian network model \cite{koller2009probabilistic} for artificial intelligent agent decision-making in uncertain and adversarial environments. It builds hierarchical relationships between individual behaviors and interactive entities and helps agents infer the following step strategy combination representing more complex behaviors to adapt to various scenarios through given the observations and the arriving data in real-time. 
    
\paragraph{Relative Needs Entropy (RNE)} Based on the \textit{Robot Needs Hierarchy} model, \textit{Trust} can be defined as the difference or distance of needs distribution between agents or groups in a specific scenario. 
Statistics regard \textit{Relative Entropy} as the similarity of high-dimensional sample sets. Here, we call it -- \textit{Relative Needs Entropy (RNE)} \cite{yang2021can}, which describes the reliability and stability of the relationships between agents in MAS cooperation.
    
\paragraph{GUT-Based Bayesian Adaptive Learning} 
By designing the appropriate \textit{Trust (RNE) Network}, building suitable perception models, estimating reasonable parameters, simplifying the corresponding \textit{GUT} model, and implementing structure learning (Fig. \ref{fig: sass_sf}), adaptive learning \cite{nikolaev2006adaptive} can efficiently optimize the group strategy fitting the specific scenarios in MAS interaction and cooperation.


\paragraph{Explore Domain} From the realistic and practical perspective, \textit{Explore Domain} \cite{yang2020game} analyzes how to organize more complex relationships and behaviors in MAS cooperation, achieving tasks with higher success probability and lower costs in adversarial environments.

\paragraph{Urban search and rescue (USAR)} Through organizing heterogeneous agents' needs for MAS cooperation in USAR, we consider the \textit{Group's Utility} (teaming needs) as the number of victims or valuable properties rescued as much as possible in a limited time \cite{yang2020needs}.

\section{Conclusion}
\label{sec:conclusions}


In the \textit{SASS}, \textit{Robot Needs Hierarchy} is the foundation. It surveys the system's utility from individual needs. Balancing the rewards between agents and groups for MAS through interaction and adaptation in cooperation optimizes the global system's utility and guarantees sustainable development for each group member, much like human society does.

As a novel DAI model, 
the planning and control govern the individual low-level \textit{safety} and \textit{basic} needs; \textit{capability} and \textit{teaming} needs are the preconditions and requirements of MAS cooperation in decision-making for achieving tasks; individuals upgrade themselves from interaction, cooperation, and adaptation in the process  for the highest level needs \textit{learning}, helping \textit{SASS} self-evolution.


\bibliographystyle{named}
\bibliography{ijcai21}

\begin{thebibliography}{}

\bibitem[\protect\citeauthoryear{Koller and
  Friedman}{2009}]{koller2009probabilistic}
Daphne Koller and Nir Friedman.
\newblock {\em Probabilistic graphical models: principles and techniques}.
\newblock MIT press, 2009.

\bibitem[\protect\citeauthoryear{Nikolaev and Iba}{2006}]{nikolaev2006adaptive}
Nikolay Nikolaev and Hitoshi Iba.
\newblock {\em Adaptive learning of polynomial networks: genetic programming,
  backpropagation and Bayesian methods}.
\newblock Springer Science \& Business Media, 2006.

\bibitem[\protect\citeauthoryear{Rizk \bgroup \em et al.\egroup
  }{2018}]{rizk2018decision}
Yara Rizk, Mariette Awad, and Edward~W Tunstel.
\newblock Decision making in multiagent systems: A survey.
\newblock {\em IEEE Transactions on Cognitive and Developmental Systems},
  10(3):514--529, 2018.

\bibitem[\protect\citeauthoryear{Rizk \bgroup \em et al.\egroup
  }{2019}]{rizk2019cooperative}
Yara Rizk, Mariette Awad, and Edward~W Tunstel.
\newblock Cooperative heterogeneous multi-robot systems: A survey.
\newblock {\em ACM Computing Surveys (CSUR)}, 52(2):1--31, 2019.

\bibitem[\protect\citeauthoryear{Stone and Veloso}{2000}]{stone2000multiagent}
Peter Stone and Manuela Veloso.
\newblock Multiagent systems: A survey from a machine learning perspective.
\newblock {\em Autonomous Robots}, 8(3):345--383, 2000.

\bibitem[\protect\citeauthoryear{Wang and Yeung}{2020}]{wang2020survey}
Hao Wang and Dit-Yan Yeung.
\newblock A survey on bayesian deep learning.
\newblock {\em ACM Computing Surveys (CSUR)}, 53(5):1--37, 2020.

\bibitem[\protect\citeauthoryear{Wooldridge}{2009}]{wooldridge2009introduction}
Michael Wooldridge.
\newblock {\em An introduction to multiagent systems}.
\newblock John Wiley \& Sons, 2009.

\bibitem[\protect\citeauthoryear{Yang and Parasuraman}{2020a}]{yang2020game}
Qin Yang and Ramviyas Parasuraman.
\newblock A game-theoretic utility network for cooperative multi-agent
  decisions in adversarial environments.
\newblock {\em arXiv preprint arXiv:2004.10950}, 2020.

\bibitem[\protect\citeauthoryear{Yang and
  Parasuraman}{2020b}]{yang2020hierarchical}
Qin Yang and Ramviyas Parasuraman.
\newblock Hierarchical needs based self-adaptive framework for cooperative
  multi-robot system.
\newblock In {\em 2020 IEEE International Conference on Systems, Man, and
  Cybernetics (SMC)}, pages 2991--2998. IEEE, 2020.

\bibitem[\protect\citeauthoryear{Yang and Parasuraman}{2020c}]{yang2020needs}
Qin Yang and Ramviyas Parasuraman.
\newblock Needs-driven heterogeneous multi-robot cooperation in rescue
  missions.
\newblock In {\em 2020 IEEE International Symposium on Safety, Security, and
  Rescue Robotics (SSRR)}, pages 252--259. IEEE, 2020.

\bibitem[\protect\citeauthoryear{Yang and Parasuraman}{2021}]{yang2021can}
Qin Yang and Ramviyas Parasuraman.
\newblock How can robots trust each other? a relative needs entropy based trust
  assessment models.
\newblock {\em arXiv preprint arXiv:2105.07443}, 2021.

\bibitem[\protect\citeauthoryear{Yang \bgroup \em et al.\egroup
  }{2019}]{mrs2019}
Qin Yang, Zhiwei Luo, Wenzhan Song, and Ramviyas Parasuraman.
\newblock {Self-Reactive Planning of Multi-Robots with Dynamic Task
  Assignments}.
\newblock In {\em IEEE International Symposium on Multi-Robot and Multi-Agent
  Systems (MRS) 2019}, 2019.
\newblock Extended Abstract.

\end{thebibliography}

\end{document}